# Warp and Weft Wiring method for rapid, modifiable, self-aligned, and bonding-free fabrication of multi electrodes microfluidic sensors

Ali Olyanasab[1], Zahra Meskar[2], Mohsen Annabestani[3*], Ali Mousavi Shaegh[4], Mehdi Fardmanesh[5*]

[1] Department of Electrical Engineering, Sharif University of Technology, Tehran, Iran
Email: olyanasab.a@gmail.com

[2] Department of Electrical Engineering, Sharif University of Technology, Tehran, Iran
Email: nadia.meskar@yahoo.com

[3] Department of Electrical Engineering, Sharif University of Technology, Tehran, Iran
Email: annabestany@gmail.com

[4] Mashhad University of Medical Sciences, Mashhad, Iran
Email: a.m.shayegh@gmail.com

[5] Department of Electrical Engineering, Sharif University of Technology, Tehran, Iran
Email: fardmanesh@sharif.edu

**Abstract**

The need for rapid fabrication of microfluidic devices has become increasingly critical as microfluidics become part of biomedical sensors. Using Warp and Weft Wiring (WWW) of copper wires, this paper presents a novel low-cost method for rapid, self-aligned, bonding-free, and modifiable fabrication of multi-electrodes microfluidic sensors. All the proposed features are promising and highly recommended for the development of Point-of-Care Tests (POCTs), while most of the conventional methods have low chances of coming out of the research labs and play no role in POCTs development. To have an experimental proof of concept, the proposed chip was fabricated and then tested with two sets of experiments that showed the potential applications of water quality management, hygiene, biomedical impedance measurement, cell analysis, flow cytometry, etc.

**Introduction**

Microfluidic devices have become increasingly popular in applications. Some of the trending applications of microfluidic devices are Probing small samples[1]–[5], Diagnosing [6]–[9], Therapy [3], [10], [11], Food safety [12]–[14], and production and formulation of materials [15]–[20].

A wide range of microfluidic chip fabrication methods have been developed as a result of increased applications in microfluidics. For example, lithography [21], Soft lithography [22], Injection molding [23], Bonding [24], Laser engraving [25], [26], Inkjet printing [27], [28], 3D printing [29], [30], and Micromachining [31] can be mentioned which are popular among researchers.

There have been studies on rapid prototyping methods of microfluidic devices in recent years. Thompson et al. describe a technique for fabricating microfluidic devices with complex multilayer architectures using a laser printer, a co2 laser cutter, an office laminator, and common overhead transparencies as a printable substrate via a laser print, cut, and laminate (PCL) [32]. Kaigala et al. demonstrated a rapid and inexpensive approach to fabricating high-resolution PDMS-based microfluidic devices [33]. Annabestani et al. presented a new method to fabricate modifiable chips made of ethylene-vinyl acetate (EVA) quickly [34]. Mukherjee et al. demonstrated the fabrication of soft photolithography masters using lamination of ADEX dry film as an alternative to SU-8 resist masters [35]. Nascak et al. proposed a fabrication process of reusable microscope-compatible microfluidic devices. In order to implement the method, some commonly available equipment, such as a 3D printer, a hot plate, etc., must be available. [36]. Zhang presented a new method for rapidly prototyping a microfluidic device on a piezoelectric substrate [37]. Lai et al. demonstrated a novel method for rapid prototyping microfluidic devices using inkjet pattern-guided liquid templates on superhydrophobic substrates [38]. Qin et al. reported a photolithography-free method for rapidly producing PDMS devices within 30 min [39]. Ostmann et al. proposed a masking method for SU8-based soft lithography utilizing a simple optical setup using transparencies printed by a standard office printer [40]. Most methods reviewed cannot be modified, have planar electrodes, are expensive, take a considerable amount of time to construct, and were developed to fabricate with a single material. Our method presents a low-cost, rapid fabrication method for modifiable microfluidic devices with 3D electrodes and a completely uniform channel that can be made with a wide range of materials.

**Proposed method**

In the majority of previous studies, fabrication of a microfluidic device with a channel thickness of less than 100um has required high-resolution apparatus and complex processes. Making microfluidic devices with molds is one method of fabrication. Any fabrication using a single mold would result in the same uniform and persistent chip due to its solid and fixed nature. If we could propose a simple idea to fabricate the microfluidic channel with something like an inner mold, we would have a microfluidic device with a uniform structure and a uniform channel and electrodes; meanwhile, the fabrication time would be short. A possible solution would be printing the mold for the microfluidic chip and channel using two-photon polymerization. While two-photon polymerization is relatively slow (~0.1mm3/hour), its hardware is vastly expensive. In addition, the shape and size of the microfluidic channel require a challenging bonding process. We propose using a 3D printed structure as the mold and wires as the inner molds to fabricate the whole microfluidic chip simultaneously with no bonding requirement. It is possible to self-align the wires by fixing the head and tail of the wires since there is only one straight path between two fixed points. **(Figure 1)** Shows the designed structure for the proposed method, which we call Warp and Weft Wiring method for microfluidic chip fabrication or WWW. The main structure is 3D printed, serves as the mold for the microfluidic chip, and is reusable. Therefore, the printing time of the main structure will not affect the fabrication time of the microfluidic chip.

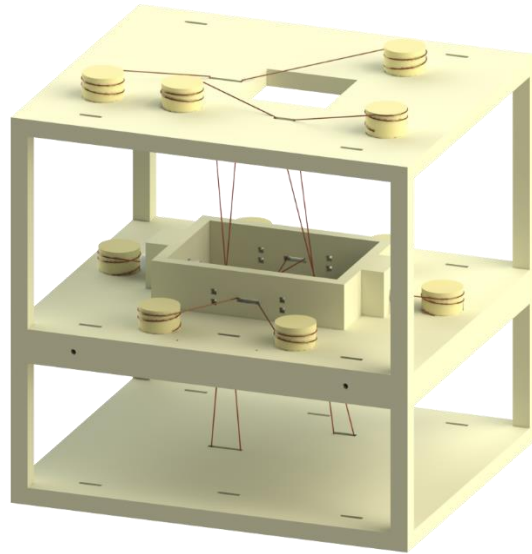

**Figure 1.** The 3D structure of the WWW method for rapid fabrication of microfluidic chips.

The proposed microfluidic chip could be manufactured using wires of different materials and thicknesses (from 5um to 200um), allowing for easy modification of the electrodes and channel. A further advantage of this method is that one can fabricate a microfluidic chip with 3D electrodes at any desired angle, which yields more details about the fluid flowing through the channel.

**Fabrication Procedure and Hardware apparatus**

The fabrication steps of the proposed method are shown in **(Figure 2)**. First, the designed 3D structure was printed using an FDM 3D printer. The printing material for the printer was Acrylonitrile Butadiene Styrene (ABS), a common thermoplastic polymer. Then the 3D printed cylinders and soft silicone tubes were placed. Afterward, a rod of Ecoflex$^{TM}$ was made by a syringe and cured using a hot air gun. After that, the mold was sealed using hot glue and poured with Epoxy resin. Finally, the electrode wires were detached from cylinders, the channel wire was pulled out, and the fabricated chip was separated from the mold. Another advantage of this approach is that the contact of electrodes and the channel are diagnosable and maintainable. Because the electrodes and the channel wire are conductive, a simple multimeter could be used to assure the contact of electrodes and the channel wire in **(Figure 2 (e))** step.

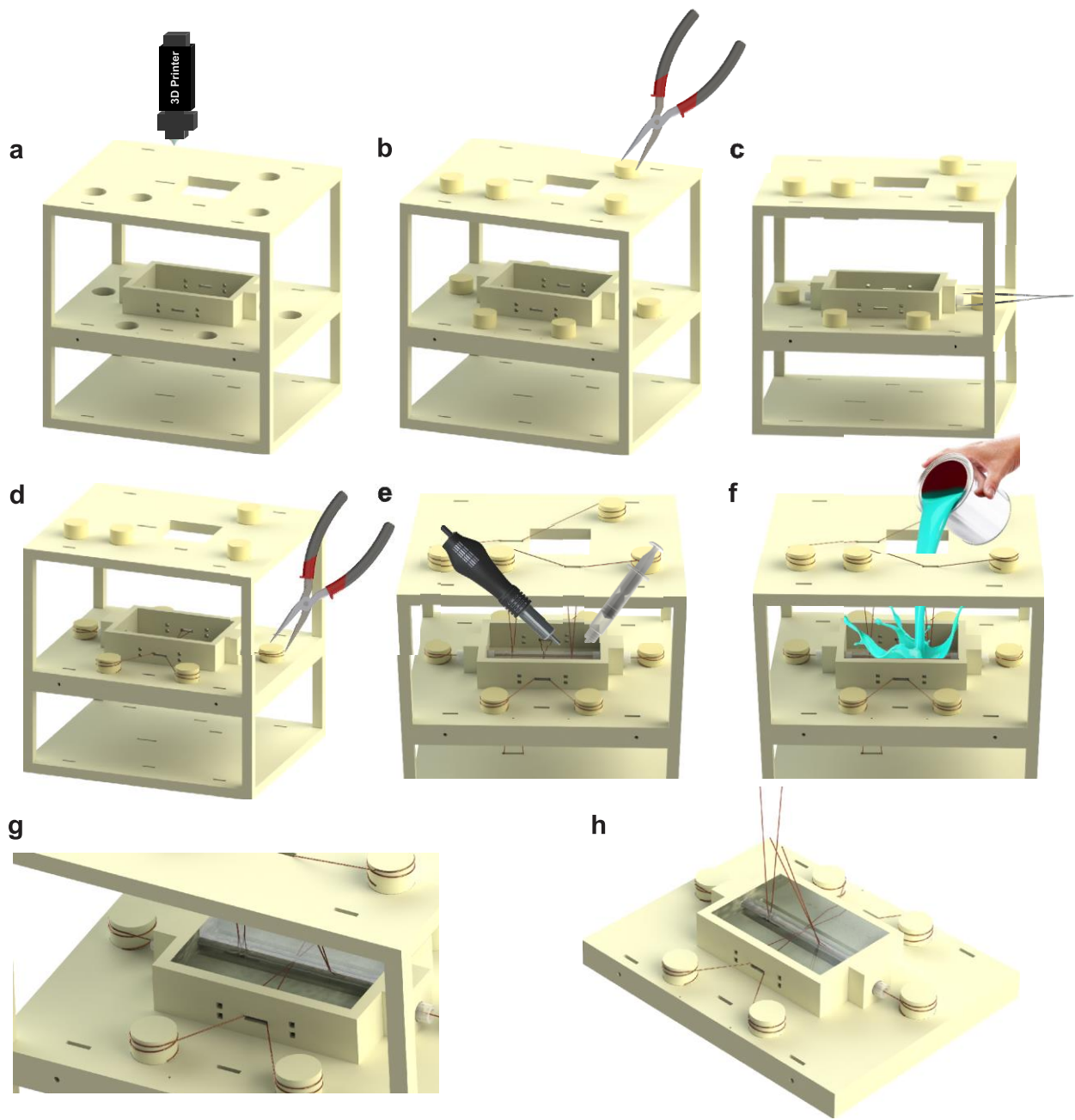

**Figure 2.** Fabrication procedure. (a) printing the structure using a 3D printer (b) fixing the cylinders (c) placing the soft silicone tube (d) wiring (e) making a rode of Ecoflex$^{TM}$ 00-30 using a syringe and curing the rod using a hot air gun (f) pouring the mold with Epoxy resin (g) putting the chip on the hot plate and wait to cure (h) detaching the 3D electrode wires from cylinders, disconnecting the inner and outer mold.

(**Figure 3**) shows the fabricated chip made of Ecoflex<sup>TM</sup> 00-30 and Epoxy resin with three pairs of electrodes; A pair of planar electrodes, a pair of perpendicular electrodes, and a pair of 3D electrodes with a 30° angle with the surface. The number and the angle of the electrodes could be adjusted for customized purposes. Also, the materials used in the proposed method could be alternated with other materials and silicone rubbers such as Dragon Skin<sup>TM</sup>, PDMS, etc. The fabricated chip is translucent, which makes it suitable for optical analysis.

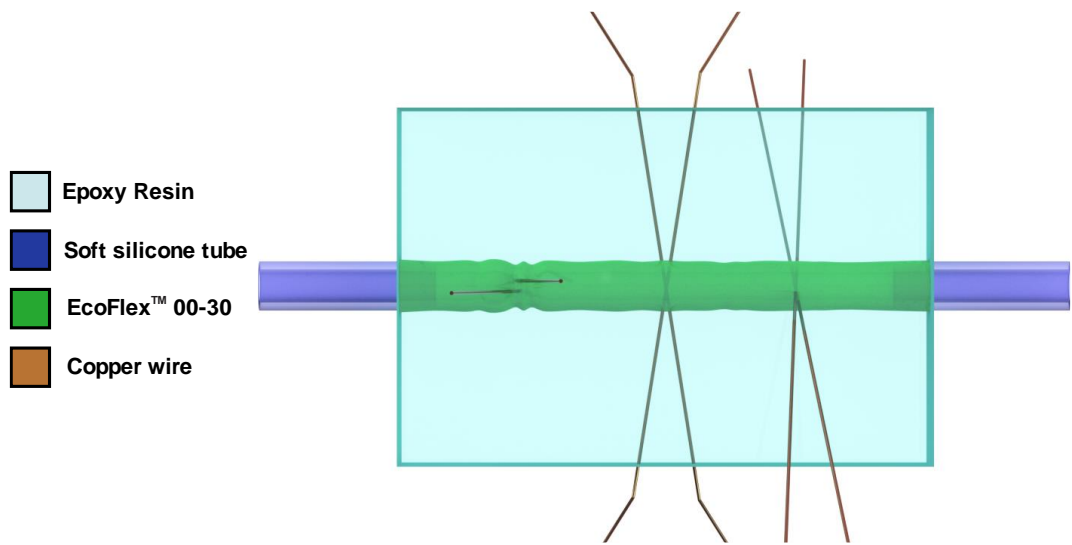

**Figure 3.** Schematic of the fabricated microfluidic chip with three pairs of 3D electrodes from the top view.

**Experiments and hardware apparatus**

Two experiments were conducted to evaluate the proposed microfluidic chip and assure its potential applications. First, a set of fluids was prepared, and their impedance was measured using the fabricated chip. (**Figure 4**) shows the experimental setup for impedance measurement of different fluids.

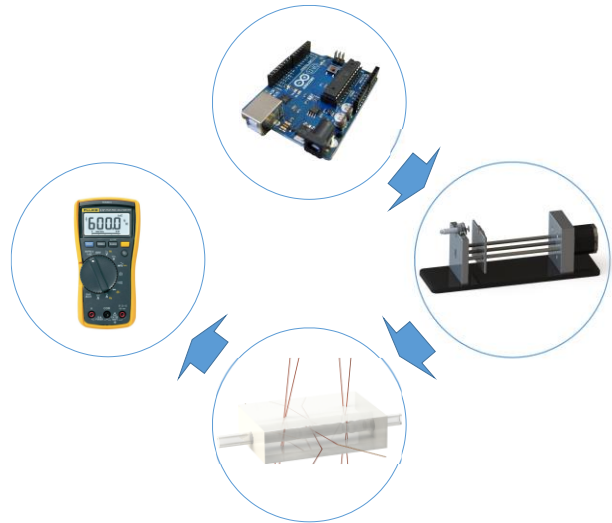

**Figure 4.** Experimental setup for impedance measurement of fluids

First, the syringe was filled with different fluids and placed in the pump. A commercial LCR meter was used to determine the impedance of several fluids at the frequency of 1kHz, which will be illustrated and discussed in the following section.

In the second experiment, a microfluidic system was developed to evaluate the proposed sensor for detecting Polystyrene beads. **(Figure 5)** illustrates the schematics of the microfluidic system used to test the fabricated microfluidic chip.

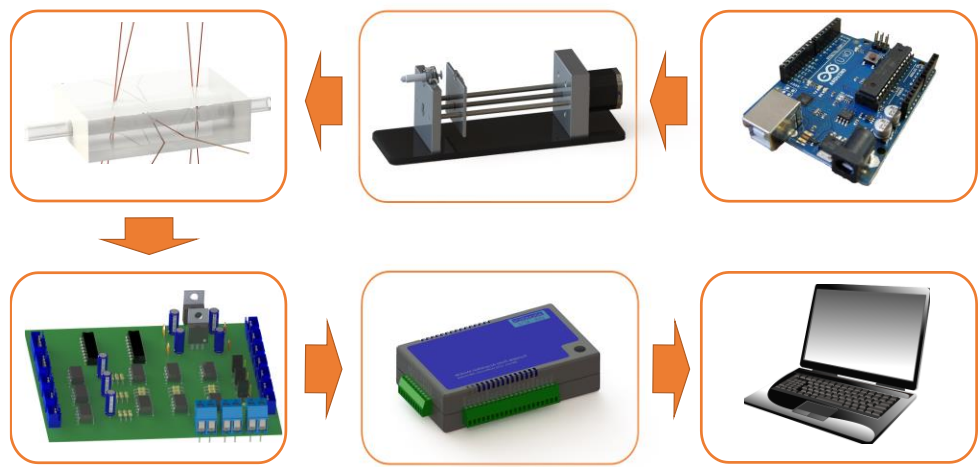

**Figure 5.** Experimental setup for evaluation of the fabricated chip

As shown in the figure, an Arduino Uno was used to control the syringe pump. The syringe pump was implemented using 3D printed parts and a Stepper motor. The syringe was connected to the fabricated chip using the soft tubes implemented in the microfluidic chip. The chip's electrodes were connected to the electrical circuit, the circuit's output was connected to the DAQ, and the DAQ was connected to the PC. The electrical circuit was a customized low-noise lock-in amplifier for fast impedance measurement of the microfluidic channel. Advantech USB-4716 was used as DAQ, which has a 16-bit resolution with a sampling rate of 200kSa/s.

First, a diluted solution of PBS/BSA with polystyrene beads with a thickness of 15um was prepared. Then the syringe was filled with the prepared solution and placed on the syringe pump. Afterward, the experiment started with a flow rate of 1uL/min to determine the presence of the PS beads in the microfluidic channel. The result of this experiment is shown and discussed in the next section.

**Results and discussion**

The first experiment's results are shown in **(Table 1)** demonstrating the functionality of the fabricated sensor for hygiene, water quality management, biomedical impedance measurement, etc. In order to utilize the microfluidic chip in each of the above applications, the microfluidic system should be calibrated, which is the usual procedure for microfluidic biosensors.

**Table 1:** Results of impedance measurement with different fluids

| Fluid | Impedance @ 1kHz |
|---|---|
| Dry channel | Open load |
| Drinking water | 356kΩ |
| Salt solution | 18.57kΩ |
| Sugar solution | 286kΩ |
| Commercial hand synthesizer (70% Ethanol, water, etc.) | 1.05MΩ |

**(Figure 6)** represents the frequency spectra of the output signal, calculated using the Fast-Fourier Transform of the output signal. It is apparent that the electrical system has low harmonic distortion. The DC signal is related to the impedance of the microfluidic channel. The 2kHz signal, which is the second harmonic of the input signal, could be easily eliminated using a digital notch filter. Additionally, the output signal's noise floor was calculated, which is about -100dB over 100kHz.

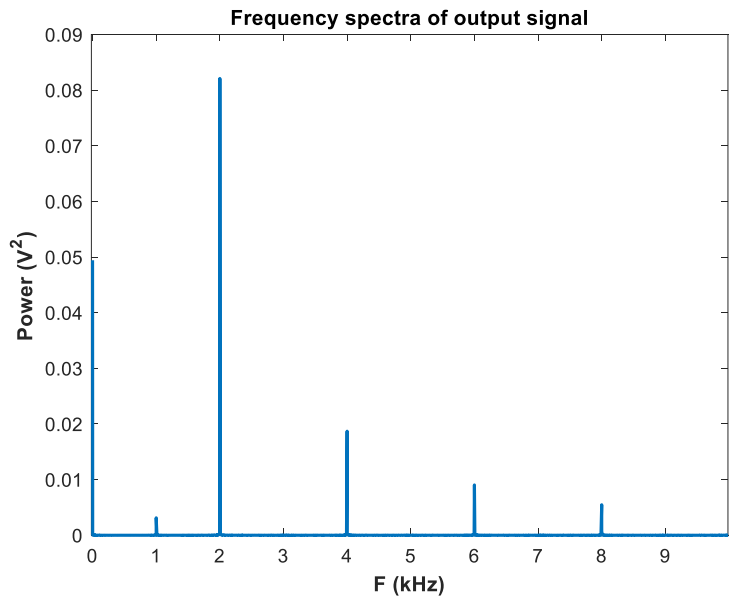

**Figure 6.** Frequency spectra of the output signal

**(Figure 7)** illustrates the change of the output signal due to the change of the fluid in the microfluidic channel. The first fluid was drinking water and the second fluid was PBS/BSA solution. The result indicates the microfluidic system's relatively fast response to the fluid change.

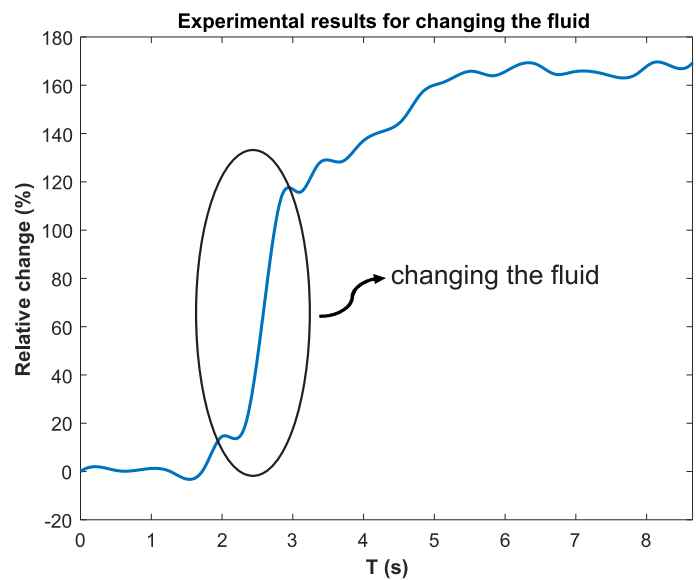

**Figure 7.** The experimental results of changing the fluid of the microfluidic channel from drinking water to PBS/BSA solution.

**(Figure 8 (a))** shows the results of the second experiment. Observations of the PS beads in the microfluidic channel are evident from the electrical signal peaks. Also, a durability test was done with PBS/BSA solution without any PS beads to examine the output signal's robustness and stability, shown in **(Figure 8 (b))**, representing good stability in the signal output. As the results indicate, the fabricated microfluidic chip demonstrates its ability to analyze cells and perform flow cytometry. To determine the precise thickness of PS beads in a multiparticle manner, we will use machine learning and signal processing algorithms in future research.

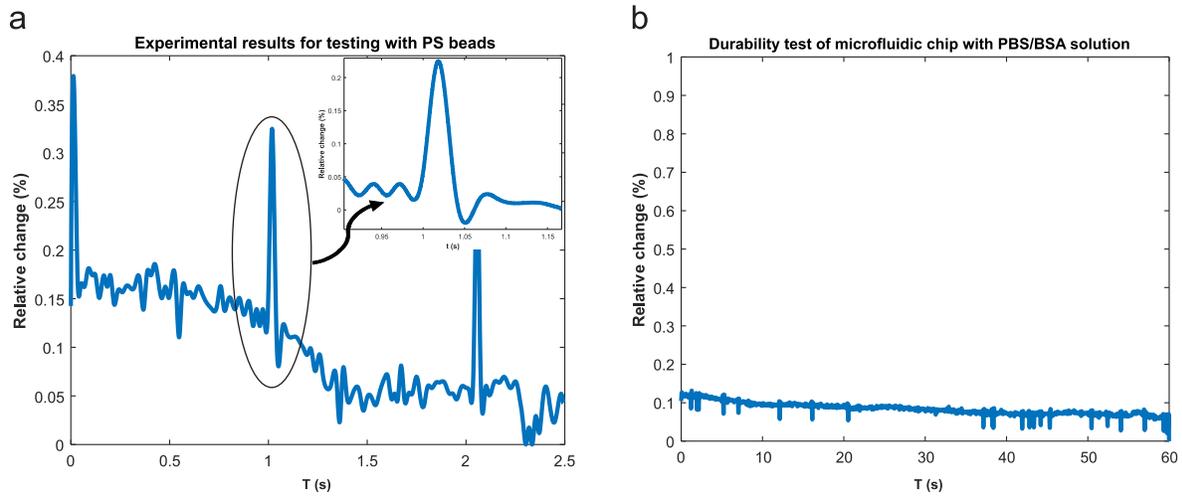

**Figure 8.** Experimental results of testing microfluidic chip with (a) PBS/BSA solution and 15um PS beads. (b) PBS/BSA solution without PS beads

## Conclusion

Our paper presents a novel approach for rapid fabrication of low-cost, self-aligned, modifiable, and bonding free multi-electrodes microfluidic sensors. An experiment was conducted to determine the impedance of various fluids using the fabricated chip. Results indicated that the fabricated chip was functional for hygiene, water quality management, and biomedical impedance measurement. Then a second experiment was conducted using PBS/BSA solution with PS beads with a thickness of 15um. The results demonstrated the functionality of the proposed microfluidic chip for cell analysis and flow cytometry applications.